\documentclass[oneside,letterpaper]{amsart}
\usepackage[latin1]{inputenc}
\usepackage{amssymb}
\usepackage{geometry}
\usepackage{float}
\usepackage{graphicx}
\usepackage{appendix}
\usepackage{amsrefs}

\makeatletter
\theoremstyle{plain}
\newtheorem{thm}{Theorem}[section]
\newtheorem{conj}{Conjecture}[section]
\theoremstyle{plain}
\newtheorem{fact}[thm]{Fact}
\newtheorem*{remark*}{Remark}
\newtheorem*{acknowledgement*}{Acknowledgement}
\theoremstyle{plain}

\makeatother
\begin{document}

\title{A Gauge-independent Mechanism for Confinement and Mass Gap: Part II --- $G=SU(2)$ and $D=3$}

\author{J. Wade Cherrington$^{1}$}

\maketitle
\vspace*{-10pt}

\begin{center}{\small $^{1}$Department of Mathematics, University
of Western Ontario, London, Ontario, Canada}\par\end{center}{\small \par}

\maketitle

\begin{abstract}
We apply to the case of gauge group $G=SU(2)$ in three dimensions a recently proposed
gauge-independent mechanism for confinement that is based on a particular
form of the dual spin foam framework for lattice gauge theory. Explicit formulae for interaction
factors and their asymptotics are introduced and their behavior in different sectors of the theory are
identified and analyzed.  We arrive at several elementary properties of the dual 
theory that represent one scenario by which confinement may be realized at weak coupling.
We conclude with an outlook for further development of this approach.
\end{abstract}

\section{Introduction}
The present work provides a detailed application of a new framework for investigating the confinement
of static charges and mass gap in Yang-Mills theory to the case of gauge group $G=SU(2)$ and space-time 
dimension $D=3$. This framework was presented in~\cite{Conf1} and is based on an 
exact duality between conventional Yang-Mills in the lattice regularization and a lattice spin foam model.
The case of $G=SU(2)$ in three dimensions is particularly attractive as a first testing ground for the
general case, as explicit vertex amplitudes are known and possess a particularly tractable
form~\cite{AS,ACSM,AGMS91}.  At the same time, this case is rich enough to exhibit in a non-trivial 
fashion the various features introduced in~\cite{Conf1}. A previous application of spin foams to the
confinement problem for this case has also been made based on analogy with the dual photon and monopole picture
of confinement in $U(1)$ theories~\cite{ConradyGluons}; the present work also uses the spin foam
transformation but proceeds in a different direction.

Broadly speaking, the results we obtain here can be summarized as follows. Using an exact duality in terms
of spin foams we show explicitly from first principles how an area law behavior for the expectation
values of Wilson loop observables may be obtained in the weak coupling limit.  While short of a
rigorous proof, we can clearly identify properties that would be required
to provide such a proof --- they are given as conjectures.
Thus, from the perspective of finding a rigorous proof of confinement, what we have done is present
a new framework, within which can be formed a reduction of the confinement problem into several
``smaller'' conjectures that are elementary propositions regarding the spin foam model.
In doing so, we hope to invite attention to this new perspective that stands apart from many past approaches
in offering an understanding of confinement and mass gap that is manifestly gauge-invariant.

The structure of the paper is as follows.  In Section 2, we recall the spin foam expansion of the 
expectation value of a Wilson loop observable.  Applying the general framework of~\cite{Conf1},
the worldsheet-background decomposition and interaction factors are found explicitly. In Section 3, we
analyze the interaction factors in more detail, and in particular introduce an asymptotic formula for 
the vertex interaction $I_{V}(f)$ that holds in the large spin limit.  Applying asymptotic 
estimates for the vertex interaction, we find that for the class of self-avoiding worldsheets,
one can show that $I_{V} \rightarrow 1$ in the weak coupling limit --- a property we shall refer to 
as ``vertex decoupling'', a special case of the vertex stability property introduced in~\cite{Conf1}.
In Section 4, we consider different sectors of the full theory according to the framework
described in~\cite{Conf1}.  In the case of self-avoiding worldsheets interacting with asymptotic
background, we show how area-law behavior is particularly transparent due to vertex decoupling.  
The remaining sectors of the theory are also discussed and specific properties are 
identified as necessary conditions for confinement. In Section 5 we give formal statements of 
these required properties in a form suitable for pursuing via analytic methods or through
numerical method such as the recently developed non-abelian dual simulation algorithms~\cite{CCK}. 
Section 6 concludes with an assessment of what has been shown and discusses ways in which the 
results may be strengthened.

\section{The lattice spin foam form of the Wilson loop observable}
We start by introducing the passage from the conventional to spin foam form of LGT,
and review the worldsheet-background formulation of the Wilson loop observable that
was introduced in~\cite{Conf1}.

\subsection{The spin foam dual}
As reviewed in~\cite{Conf1} and references therein, the spin foam transformation of lattice Yang-Mills theory
can be summarized as follows:
\begin{eqnarray}\label{dualities}
Z_{V} &=& \int \left( \prod_{e\in E}dg_{e} \right) e^{-\sum_{p\in P}S(g_{p})} \\ \nonumber
&=&\int \left( \prod_{e\in E}dg_{e} \right) \prod_{p\in P}\left(\sum_{j_{p}=\frac{1}{2}}^{\infty}
c_{j_{p}}\chi_{j_{p}}(g_{p})\right)
=\sum_{\{ j_{p}\}}^{\infty}\int\prod_{e\in E}dg_{e} \prod_{p\in P}c_{j_{p}}\chi_{j_{p}}(g_{p}) \\ \nonumber
&=&\sum_{f\in F_{V}}\prod_{p \in P}A_{P}(f(p))\prod_{ e \in E}A_{E}(f(e))\prod_{v \in V}A_{V}(f(v)),
\nonumber
\label{eq:ZVac}
\end{eqnarray}
where $g_{p}$ is the holonomy consisting of a product of group valued variables and their
inverses along the edges of $p$. In the present case of $G=SU(2)$ and $D=3$, we have
\begin{equation}\label{plaq_ampl}
A_{P}(f,p) = c_{j_{p}} = (2j_{p}+1)e^{-\frac{2}{\beta}j_{p}(j_{p}+1)},\end{equation}
\begin{equation}
A_{E}(f,e) = \begin{matrix} \includegraphics[scale=0.8]{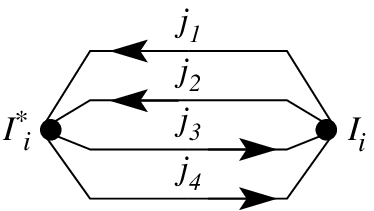} \end{matrix},
\label{eq:edgeampl}
\end{equation}
and
\begin{equation}
A_{V}(f,v) = \begin{matrix} \includegraphics[scale=0.6]{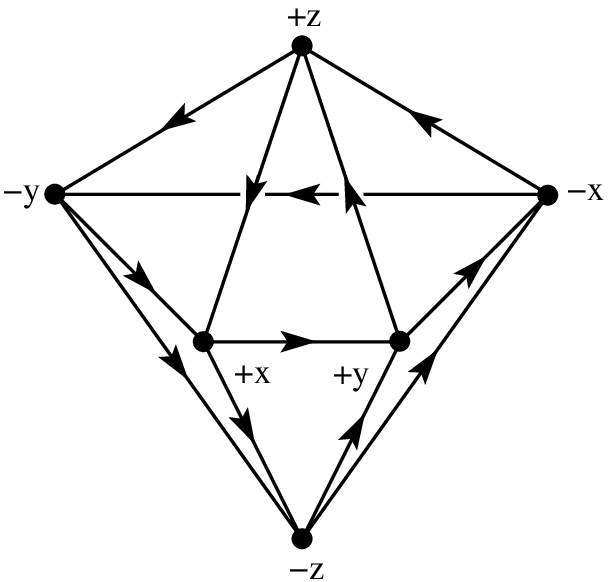} \end{matrix}.
\end{equation}
The first line of~(\ref{dualities}) is the conventional lattice regularization of a gauge theory using 
a discretized action $S(g_{p})$. The second line represents the character expansion of the action that
has often been used in strong coupling expansions for lattice gauge theory~\cite{DZ1983}.  The final line is 
based on a lattice spin foam duality, which can be understood as a refinement of the character expansion
method; Consistent with the formalism introduced in~\cite{Conf1}, for a given spin foam 
$f$ the notation $f(p)$, $f(e)$, and $f(v)$ denotes the set of irrep and intertwiner variables local to a plaquette, edge, and vertex respectively.
A general description of this transformation can be found in~\cite{OecklPfeiffer} while
details of the transformation for our present case is discussed in~\cite{AS,ACSM,AGMS91,CCK,ConradyPolyakov}.

The configuration space of the spin foam model is $F_{V}$, the set of admissible spin foams --- all colorings 
of the plaquettes and edges by unitary irreps of $SU(2)$ satisfying certain local constraints. Admissibility 
constraints are necessary but not sufficient conditions for the edge and vertex amplitudes (and hence overall
amplitude) to be non-vanishing; a discussion can be found for example in~\cite{CCK}. 

To provide a concrete model, we assume an intertwiner basis has been chosen, in which case the above diagrams 
become spin network evaluations that can be efficiently computed on the computer, as described in~\cite{CCK}. 
Here and throughout we use the character expansion coefficients $c_{j_{p}}$ from the heat kernel action~\cite{Menotti}.

A commonly used order parameter for confinement (and the one we shall use in what follows) is the expectation 
value of a Wilson loop observable. This expectation value is defined as follows:
\begin{equation}
\left\langle O_{\Gamma_{w}}\right\rangle \equiv \frac{\int \left( \prod_{e\in E}dg_{e} \right) \,
\text{Tr} \left( \prod_{e\in\Gamma}D_{\frac{1}{2}}(g_{e}) \right) e^{-\sum_{p\in P}S(g_{p})}}
{\int \left( \prod_{e\in E}dg_{e} \right)\, e^{-\sum_{p\in P}S(g_{p})}},
\label{eq:WilsonExpect1}
\end{equation}
where $\prod_{e\in\Gamma}D_{\frac{1}{2}}(g_{e})$ denotes a product of representation matrices $D_{\frac{1}{2}}(g)$
in the $\frac{1}{2}$th (fundamental) irreducible representation of $SU(2)$\footnote{Wilson loops in higher representations
can also be analyzed by the methods discussed here but won't be detailed in the present work.}.

\begin{figure}[htb]
\includegraphics[scale=0.65]{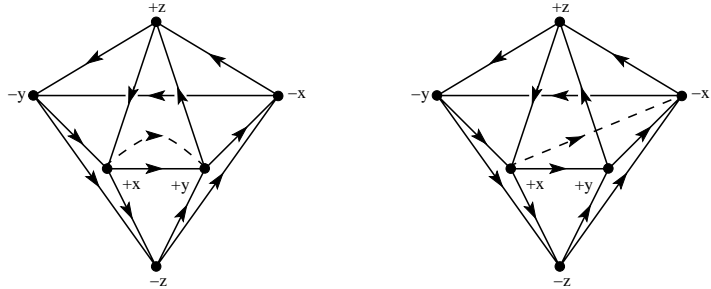}
\caption{Charged vertex amplitudes: examples of perpendicular and parallel threading.}
\label{fig:18j_charged}
\end{figure}

As shown in \cite{CherringtonFermions,ConradyPolyakov}, the presence of field insertions
associated to the Wilson loop leads to an expansion in spin foams
\begin{eqnarray}
Z_{\Gamma} &=& \int \left( \prod_{e\in E}dg_{e} \right) \, \text{Tr} \left( \prod_{e\in\Gamma}D_{\frac{1}{2}}(g_{e})
\right) e^{-\sum_{p\in P}S(g_{p})} \\ \nonumber
&=& \sum_{\{ j_{p}\}}\int \left( \prod_{e\in E} dg_{e} \right) \text{Tr}\left(\prod_{e \in \Gamma}D_{\frac{1}{2}}(g_{e}) \right)
\prod_{p\in P}c_{j_{p}}\chi_{j_{p}}(g_{p}) \\
&=& \sum_{f\in F_{\Gamma}}\prod_{p \in P}A_{P}(f(p))\prod_{e \in E}\bar{A}_{E}(f(e))\prod_{V}\bar{A}_{V}(f(v)),
\nonumber
\label{eq:ZW}
\end{eqnarray}
where the admissibility conditions of spin and intertwiner labels are changed along
the Wilson line and modified edge and vertex amplitudes and $\bar{A}_{E}$
and $\bar{A}_{V}$ are to be used.   The barred vertex amplitude has the
form given in Figure~\ref{fig:18j_charged} for edges and vertices in $\Gamma$,
with the vertex amplitude depending on whether the Wilson loop enters and leaves
in the same direction or makes a perpendicular turn at the vertex.   
For the barred edge amplitude along $\Gamma$, an insertion of a line of half-integer charge 
modifies the vacuum edge amplitude given in~(\ref{eq:edgeampl}).
For edges and vertices not in $\Gamma$, the barred amplitudes are equal to the original unbarred amplitudes.
In summary, we have rewritten the conventional form of the Wilson loop expectation value~(\ref{eq:WilsonExpect1})
as a ratio of the partition function of two different systems: 
\begin{equation}
\left\langle O_{\Gamma_{w}}\right\rangle =\frac{Z_{\Gamma}}{Z_{V}}=\frac{\sum_{f\in F_{\Gamma}}\prod_{p \in P}
A_{P}(f(p))\prod_{e \in E}\bar{A}_{E}(f(e))\prod_{v \in V}\bar{A}_{V}(f(v))}{\sum_{f\in F_{V}}
\prod_{p \in P}A_{P}(f(p))\prod_{e \in E}A_{E}(f(e))\prod_{v \in V}A_{V}(f(v))}.
\label{eq:WilsonExpect}
\end{equation}
Here we note that $Z_V$ is built of up of only vacuum configurations $F_{V}$, and $Z_{\Gamma}$ 
is built up of configurations $F_{\Gamma}$ that satisfy modified admissibility conditions 
(note that $F_{V}$ and $F_{\Gamma}$ are disjoint sets).
Since $Z_{V}$ is independent of $\Gamma$,  the dependence of the expectation value on $\Gamma$ is
controlled by $Z_{\Gamma}$ (a given lattice and fixed $\beta$).

As shown in~\cite{Conf1}, the spin foam transformation offers us a view of the Wilson loop expectation
value that shares many similarities with the strong coupling expansion but retains its validity in the continuum
limit. We describe this next.

\subsection{Definitions and decomposition of $\Gamma$-spin foams}

Let $F_{\Gamma}$ denote the set of all spin foams admissible in the presence of the given 
Wilson loop $\Gamma$ (\emph{$\Gamma$-admissible spin foams}), and let $F_{V}$ denote the 
set of spin foams admissible in the case of no external charge; we shall refer to 
these as \emph{vacuum admissible spin foams}. Finally, let $S_{\Gamma}$ denote the set of spin 
foams which can be viewed as fundamentally charged (possibly self-intersecting)
worldsheets ending on $\Gamma$; we shall refer to these as \emph{$\Gamma$-attached worldsheets}.

We next introduce an important fact that was shown in Theorem 3.1 of~\cite{Conf1}:
\begin{fact}
\textbf{Worldsheet-background decomposition.} Every element of $F_{\Gamma}$
can be expressed as the sum of a \emph{vacuum spin foam} and
a \emph{$\Gamma$-attached worldsheet}. That is, there exists a map
$\rho$ from $F_{\Gamma}$ into $F_{V}\times S_{\Gamma}$:
\begin{equation}
\rho: F_{\Gamma} \rightarrow F_{V} \times S_{\Gamma}
\end{equation}
such that the inverse map  $F_{\Gamma}$  satisfies  $\forall (f,s) \in \rho(F_{\Gamma})$
\begin{equation}
\rho^{-1}[(f,s)] = f + s,
\end{equation}
where the set $\rho(F_{\Gamma})$ is the image of the set $F_{\Gamma}$ under $\rho$ and
the  $+$ operation is the addition of half-integer labels of a configuration from $S_{\Gamma}$ at 
the corresponding plaquettes and intertwiner labels at the corresponding edges of the lattice $\kappa$.
\end{fact}

We shall refer to the $\rho$ in the above as a ``decomposition''.
Defining a decomposition is equivalent to specifying a pair of functions we shall call  $f(\gamma)$
and $s(\gamma)$ that pick out the $F_{V}$ and $S_{\Gamma}$ components of $\gamma \in F_{\Gamma}$,
respectively. 

We should emphasize here that as discussed in~\cite{Conf1}, there is no unique way of choosing
this decomposition. A given decomposition $\rho = (f(\gamma),s(\gamma))$ constitutes one
(of many possible) maps with the desired properties. 
As discussed in~\cite{Conf1}, one way of dealing with this degeneracy is to carry out unrestricted sums over 
$F_{V} \times S_{\Gamma}$, but weight every configuration by a factor accounting for
the degeneracy. This is the strategy we will apply below; others may be possible.

\subsection{The Worldsheet-background factored amplitude}
By making use of the above result, we can rewrite $Z_{\Gamma}$ as a double sum over 
vacuum admissible spin foams $F_{V}$ (which in the present context we shall
refer to as ``background'' spin foams) 
and the set of $\Gamma$-connected worldsheets $S_{\Gamma}$ as follows:
\begin{eqnarray}
Z_{\mathnormal{\Gamma}} &=& \sum_{f\in F_{\Gamma}}\prod_{P}A_{P}(f)
\prod_{E}\bar{A}_{E}(f)\prod_{V}\bar{A}_{V}(f) \\ \nonumber
&=& \sum_{f \in F_{V}} \left(\sum_{s \in S_{\Gamma}}D^{-1}(f,s)I(f,s) \right) A(f),
\end{eqnarray}
where
\begin{equation}
I(f,s)\equiv
\prod_{p \in s}\frac{A_{P}(p,s)}{A_{P}(p)}
\prod_{e \in s}\frac{\bar{A}_{E}(f(e),s)}{A_{E}(f(e))}
\prod_{v \in s}\frac{\bar{A}_{V}(f(v),s)}{A_{V}(f(v))}.
\end{equation}
Here we have expressed the expectation value of a Wilson loop $\Gamma$ in terms of a 
double sum:  a sum over worldsheets whose boundary is compatible with $\Gamma$ and a sum 
over background spin foams (from the vacuum ensemble).  
Because the sum over $S_{\Gamma}$ is unrestricted, the same foam from $F_{\Gamma}$ may come up
multiple times: we define this degeneracy to be $D(f,s)$.  We thus divide by this degeneracy factor to avoid over-counting;
see Appendix A of~\cite{Conf1} for further discussion of degeneracies.

Observe that the presence of the worldsheet modifies the background amplitude by a factor that is $\emph{localized}$
to the worldsheet. This factor we shall refer to as the $\emph{interaction factor}$;
it represents the re-weighting of the background amplitude due to
the presence of the worldsheet. 
We can now write the expectation value of the Wilson loop as an expectation value on the dual ensemble as
follows:
\begin{eqnarray}
\left< O_{\Gamma_{w}} \right>  &=& \frac{\sum_{f \in F_{V}} \left[ \sum_{s \in S_{\Gamma}} 
                    D^{-1}(f,s)I(f,s) \right ] A(f) }{\sum_{f \in F_{V}} A(f)} \\ \nonumber
                    &=& \left< \Sigma \right>_{F_{V}}.
\end{eqnarray}
where we define the $\Sigma$ observable (a function on the dual spin foam ensemble $F_{V}$) as 
the sum all worldsheets bounding $\Gamma$ of the interaction factor weighted by the inverse degeneracy factor:
\begin{equation}
\Sigma(f) \equiv \sum_{s \in S_{\Gamma}} D^{-1}(f,s)I(f,s).
\end{equation}
We next examine the interaction factor (and in particular its weak coupling limit) in more detail.

\section{Analysis of Interaction Factors}
In this section we take a closer look at the interaction factors coming from the plaquette, edge,
and vertex amplitudes.  In particular, we shall present the large spin behavior of these quantities
that play a decisive role at weak coupling.

\subsection{Plaquette interaction} In what follows we will use a ``twice-spins'' convention of using
integers rather than half-integers to index irreducible representations. With this convention,
the local plaquette amplitude~(\ref{plaq_ampl}) becomes:
\begin{equation}
A_{P}(f,p) = c_{j_{p}}(\beta) = (j_{p}+1)e^{-\frac{1}{2 \beta}j_{p}(j_{p}+2)}.
\end{equation}
By inspection of the plaquette amplitude, it is immediately clear that one can write down the 
plaquette dependent part of the interaction factor in closed form:
\begin{eqnarray}
I_{P}(f,s)&=&\frac{A_{P}(f,s)}{A_{P}(f)}=\prod_{p \in P} \frac{e^{-(2\beta)^{-1}(j_{p}+\delta_{p}(s))(j_{p}+2+\delta_{p}(s))}}
{e^{-(2\beta)^{-1}{j_{p}(j_{p}+2)}}} \left(\frac{j_{p}+1+\delta_{p}(s)}{j_{p}+1}\right) \\
&=& \prod_{p \in P}  e^{-(2\beta)^{-1}[2\delta_{p}(s)j_{f}+\delta_{p}(s)(\delta_{p}(s)+2)]}\left(\frac{j_{p}+1+\delta_{p}(s)}{j_{p}+1}\right).
\nonumber
\end{eqnarray}
where $\delta_{p}(s)$ is the change in the spin labelling plaquette $p$ due to the presence of the worldsheet $s$.
The $j_{p}$ are of course a function of the background spin foam $f$; we suppress the $f$ dependence
for compactness. Because the interaction factor for a plaquette is simply one if it is not in the worldsheet $s$,
we can rewrite the above as product of factors over the worldsheet (rather than the entire lattice) as follows:
\begin{eqnarray}\label{plaq_explicit}
I_{P}(f,s) &=&  e^{-(2\beta)^{-1}\sum_{p \in s} [2\delta_{p}(s)j_{f}+\delta_{p}(s)(\delta_{p}(s)+2)]}
\prod_{p \in s} \left(\frac{j_{p}+1+\delta_{p}(s)}{j_{p}+1}\right).
\end{eqnarray}
We discuss how the different factors in~(\ref{plaq_explicit}) behave in both the self-avoiding and self-intersecting
cases in Section 4.

\subsection{Vertex Interaction}
We now turn to the part of the interaction factor not contained in the plaquette interaction $I_{P}$.
Using a particular recoupling discussed in~\cite{CCK} and previously reported in~\cite{AS,ACSM,AGMS91},
the edge amplitudes $A_E$ and $\bar{A}_E$ can be set to unity and all the rest of the vacuum amplitude expressed as
a product of $6j$ symbols of $SU(2)$ associated to the vertices of the lattice.  Appropriate modifications are to
be made along the Wilson line $\Gamma$; we shall presently focus on the situation on the worldsheet but away\footnote{
The modified amplitudes along $\Gamma$ will provide an interaction factor that scales as the perimeter of $\Gamma$,
whereas the rest of the worldsheet will be a product of a number of interaction factors that are bounded below
by the area of $\Gamma$, and thus more relevant for confining effects.
}
from $\Gamma$.
With this choice of recoupling, the interaction associated with the vertex can be written as
 \begin{eqnarray}\label{6j_args}
I_{V}(f,s)&=&\prod_{V}\frac{\bar{A}_{V}(f,s)}{A_{V}(f)}\prod_{E}\frac{\bar{A}_{E}(f,s)}{A_{E}(f)}
= \prod_{V}\frac{\bar{A}_{V}(f,s)}{A_{V}(f)} \\ 
&=&\prod_{i=1}^{5}\frac{\left\{ \begin{array}{ccc}
a_{i}(v,f)+\delta_{a}^{i}(v,s) & b_{i}(v,f)+\delta_{b}^{i}(v,s) & c_{i}(v,f)+\delta_{c}^{i}(v,s)\\
d_{i}(v,f)+\delta_{d}^{i}(v,s) & e_{i}(v,f)+\delta_{e}^{i}(v,s) & f_{i}(v,f)+\delta_{f}^{i}(v,s)
\end{array}\right\} }
{\left\{ \begin{array}{ccc}
a_{i}(v,f) & b_{i}(v,f) & c_{i}(v,f)\\
d_{i}(v,f) & e_{i}(v,f) & f_{i}(v,f)\end{array}\right\} }.
\nonumber
\end{eqnarray}
In the above expression, the $a_{i}(v,f),\ldots,f_{i}(v,f)$ are the arguments to the product of five (indexed by $i$)
$6j$ symbols associated with the vacuum spin foam $f$ at vertex $v$.  Upon introducing a worldsheet $s$, 
some of vertices will have their arguments displaced by some amount which we shall denote
by $\delta_{a}^{i}(v,s), \ldots, \delta_{f}^{i}(v,s)$.

\subsubsection{Vertex interactions at large spin}\label{sec:largespin}
We recall a well known approximation to the Racah-Wigner $6j$ symbol that has been conjectured to hold for large spins:
\begin{equation}
\left\{ \begin{array}{ccc}
j_{1}(v) & j_{2}(v) & j_{3}(v)\\
j_{4}(v) & j_{5}(v) & j_{6}(v)\end{array}\right\} \rightarrow\sqrt{\frac{2}{3\pi V}}\cos\left\{ \sum(j_{i}(v)+1)\frac{\theta_{i}}{2}+\frac{\pi}{4}\right\} 
\end{equation}
where 
\begin{equation}
V=V(j_{1},j_{2},j_{3},j_{4},j_{5},j_{6})=\sqrt{\frac{1}{288}}\left|\begin{array}{ccccc}
0 & 1 & 1 & 1 & 1\\
1 & 0 & j_{3}^{2} & j_{5}^{2} & j_{4}^{2}\\
1 & j_{3}^{2} & 0 & j_{1}^{2} & j_{6}^{2}\\
1 & j_{5}^{2} & j_{1}^{2} & 0 & j_{2}^{2}\\
1 & j_{4}^{2} & j_{6}^{2} & j_{2}^{2} & 0\end{array}\right|^{\frac{1}{2}}
\end{equation}
is the Cayley-Menger expression for the volume of the associated tetrahedron, which has side lengths given by the spin labels;
the determinant that appears is the \emph{Cayley-Menger determinant}.
The formula assumes the limit is taken through arguments satisfying the Euclidean condition (the tetrahedron can be isometrically
embedded in flat Euclidean space, a condition equivalent to the Cayley-Menger determinant has positive sign).

Although suggested by Ponzano and Regge as early as~\cite{PR68}, to the authors 
knowledge, the above asymptotic formula has been proven to hold rigorously only recently~\cite{JR99,JR02,GVdW09} and only in
the specific case of a rescaling limit in which the arguments of some admissible $6j$ symbol are all scaled by the same 
integer $k$. In this case of rescaling, it was also shown that the non-Euclidean cases decay exponentially~\cite{JR02,GVdW09}.
Below we shall assume that the oscillatory or closely related exponential (a good discussion can be found in~\cite{LJ09})
form of the Ponzano-Regge formula holds for any $6j$ symbol containing arguments of sufficiently large spin.
This is in the spirit of the original conjecture but a slightly stronger result (stated in Section 5) than that of rescaling 
limits proven to to date. To see this, consider that at any spin scale there are tetrahedra with very long side lengths
(high spin) which are not the uniform rescaling of any tetrahedra of smaller side lengths.

Consider $I_{V}^{v}$, which we shall define to be the contribution from a single vertex in the asymptotic limit:
\begin{equation}
I_{V}^{v}(f,s)=\frac{\sqrt{\frac{2}{3\pi V'}}\cos\left[ \sum(j_{i}(v)+1+\delta_{i}(v,s))\frac{{\theta'}_{i}}{2}+\frac{\pi}{4}\right] }{\sqrt{\frac{2}{3\pi V}}\cos\left[ \sum(j_{i}(v)+1)\frac{\theta_{i}}{2}+\frac{\pi}{4}\right] }.
\end{equation}
Here $V'$ and $\theta_{i}'$ are the volume and set of dihedral angles associated with the tetrahedron whose side lengths
have been displaced by the $\delta_{i}$.
We now consider the perturbed asymptotic formula for the $6j$ in more detail:
\begin{eqnarray}\label{eqn:vert_perturb}
&&\sqrt{\frac{2}{3\pi V'}}\cos\left[ \sum_{i}(j_{i}(v)+1+\delta_{i}(v,s))\frac{{\theta'}_{i}}{2}+\frac{\pi}{4}\right] \\
\nonumber
&&= \sqrt{\frac{2}{3\pi V'}}\cos\left[ \sum_{i}(j_{i}(v)+1+\delta_{i}(v,s))\frac{\theta_{i}+\delta\theta_{i}}{2}+\frac{\pi}{4}\right]  \\
\nonumber
&&= \sqrt{\frac{2}{3\pi V'}}\cos\left[ \sum_{i}(j_{i}(v)+1)\frac{\theta_{i}}{2}+(1+\delta_{i}(v,s))\frac{\delta\theta_{i}}{2}+j_{i}(v)\frac{\delta\theta_{i}}{2}+\frac{\pi}{4}\right] \\
\nonumber
&&=  \sqrt{\frac{2}{3\pi V'}}\cos\left[ \sum_{i} \left[ (j_{i}(v)+1)\frac{\theta_{i}}{2}+\delta\theta_{i} 
+ \frac{1}{2} j_{i}(v)\delta\theta_{i} \right] +\frac{\pi}{4} \right]. \\
\nonumber
\end{eqnarray}
For concreteness and to motivate more general conjectures, consider the case of self-avoiding worldsheets, in which
$\delta = 1$ (or $\delta = 0$, as not all spins local to vertex will be perturbed). In this case, changes in 
the dihedral angles $\delta \theta_{i}$ will go to zero in the limit of small perturbations about large ``tetrahedra''.  
This leaves the term $\sum_{i}j_{i}\delta\theta_i$. It may not be immediately clear
that this quantity is small since (although the $\delta\theta_{i}$ are getting smaller) we are studying the large $j_i$ limit. However,
when $j_{i}$ are large, if $\delta\theta_{i}$ is close to zero, we can employ the fact that 
\begin{equation}
\frac{1}{2}\sum_{i}j_{i}\delta\theta_{i}\backsim0.
\end{equation}
due to the Schlafli identity\footnote{A similar argument was used by Roberts in~\cite{JR99} in relating asymptotic formulae for tetrahedra
differing by one unit of length.}~\cite{Milnor}. We use this fact in the following: for a self-avoiding worldsheet, we 
have $\delta_{i}(v,s)=1$ and thus find
\begin{eqnarray}
\frac{ \sqrt{\frac{2}{3\pi V'}}\cos\left[ \sum_{i}(j_{i}(v)+1)\frac{\theta_{i}}{2}+(1+\delta_{i}(v,s))\frac{\delta\theta_{i}}{2}+j_{i}(v)\frac{\delta\theta_{i}}{2}+\frac{\pi}{4}\right]  }
{ \sqrt{\frac{2}{3\pi V}} \cos\left[ \sum(j_{i}(v)+1)\frac{\theta_{i}}{2}+\frac{\pi}{4}\right] } \rightarrow 1.
\end{eqnarray}
Similarly, since we are considering a ratio of $V$ with the volume $V'$ of the tetrahedron that has been perturbed by single units of spin,
\begin{equation}
\frac{V}{V'} \rightarrow 1
\end{equation}
in the limit of large $j_{i}$. Combining volume and cosine factors in the asymptotic formula allows us to conclude that
\begin{equation}\label{decoupling}
I_{V}^{v}(f,s) \rightarrow 1
\end{equation}
in the limit of large $j_{i}$ for self-avoiding worldsheets. We shall refer to this as the 
\emph{vertex decoupling} property.
While we have worked with the Euclidean case, the case of small perturbations in spin of
the non-Euclidean follows by a similar analysis.

\section{Confinement in the weak coupling limit}
We now proceed to apply the general confinement scenario proposed in~\cite{Conf1} to our present case, the
first step of which separates contributions to the expectation value coming from asymptotic and sub-asymptotic backgrounds.

\subsection{Decomposition into asymptotic and sub-asymptotic backgrounds}
Let a positive integer $j^{*}$ be given. It will be used to control the transition to asymptotic behavior for the vertex
amplitudes and interaction factors. Background foams containing any plaquette spin less than $j^{*}$ are said to be \emph{sub-asymptotic} and 
the set of such foams is denoted by $\mathcal{J}_{0}$; the complement of this set is denoted by $\mathcal{J}^{+}$ and its
elements are described as \emph{asymptotic} background foams. With these definitions, we can write the Wilson loop expectation value as follows:
\begin{eqnarray}\label{asympdecomp}
\left< O_{\Gamma_{w}} \right> &=& \left< O_{\Gamma_{w}} \right>_{0} + \left< O_{\Gamma_{w}} \right>_{+} \\ \nonumber 
&\equiv& 
Z_{V}^{-1} \sum_{f \in \mathcal{J}^{+}} \sum_{s \in S_{\Gamma} }  D^{-1} I_{P}(f,s) I_{V}(f,s)A(f),
+Z_{V}^{-1} \sum_{f \in \mathcal{J}_{0}} \sum_{s \in S_{\Gamma} }  D^{-1} I_{P}(f,s) I_{V}(f,s)A(f) 
\end{eqnarray}
where we have decomposed $\left< O_{\Gamma_{w}} \right>$ into asymptotic 
and sub-asymptotic contributions, denoted by $\left< O_{\Gamma_{w}} \right>_{+}$ and $\left< O_{\Gamma_{w}} \right>_{0}$ respectively.
This allows us to apply different approaches to the two cases; we shall discuss each in turn.

\subsection{Asymptotic backgrounds}
We decompose the first term of equation (\ref{asympdecomp}) into the set of worldsheets that are self-avoiding, $S_{\Gamma}^{1}$ and
the remaining self-intersecting worldsheets, denoted by $S_{\Gamma}^{*}$.
\begin{eqnarray}
\left< O_{\Gamma_{w}} \right>_{+} &=& Z_{V}^{-1}  \sum_{f \in \mathcal{J}^{+}} \sum_{s \in S_{\Gamma} } D^{-1} I_{P}(f,s) I_{V}(f,s)A(f) \\
\nonumber
&=& Z_{V}^{-1}  \sum_{f \in \mathcal{J}^{+}} \sum_{s \in S_{\Gamma}^{1} } D^{-1} I_{P}(f,s) I_{V}(f,s)A(f) + Z_{V}^{-1}
 \sum_{f \in \mathcal{J}^{+}} \sum_{s \in S_{\Gamma}^{*} } D^{-1}I_{P}(f,s) I_{V}(f,s)A(f).
\end{eqnarray}
We now consider the two types of worldsheets.
\subsubsection{Self-avoiding worldsheets} We start with the case of self-avoiding worldsheets, as the situation is particularly clear in this case. Consider the interaction factor:
\begin{equation}
I_{V}(f,s) = \frac{\prod_{v \in V} \prod_{i=1}^{5} \{ \vec{j}_{i,v}(f) + \vec{\delta}_{i,v}(s) \}} 
{\prod_{v \in V} \prod_{i=1}^{5} \{ \vec{j}_{i,v}(f) \} }.
\end{equation}
Based on vertex decoupling noted in Section 3 equation~(\ref{decoupling}), we apply vertex decoupling to the 
ratio $I_{V}^{i,v} = \frac{  \{ \vec{j}_{i,v}(f) + \vec{\delta}_{i,v}(s) \} } { \{ \vec{j}_{v}(f) \} }$ and find that it is 
bounded by $1+\epsilon(j^{*})$, where $\epsilon \rightarrow 0$ as $j^{*} \rightarrow \infty$. 

We next recall from  Section 3 the form of the plaquette interaction amplitude $I_{P}$:
\begin{eqnarray}\label{plaqamp}
I_{P}(f,s) &=& \prod_{p \in P(s)} e^{-(2\beta)^{-1}[2\delta_{p}(s)j_{p}+\delta_{p}(s)
(\delta_{p}(s)+2)]}\left(\frac{j_{p}+1+\delta_{p}(s)}{j_{p}+1}\right).
\end{eqnarray}
As we are in the case of self-avoiding worldsheets, we have $\delta_{p}(s)=1$ for all plaquettes in $s$ and can simplify
to 
\begin{eqnarray}
I_{P}(f,s) &=& \prod_{p \in s} e^{-\beta^{-1}(j_{f}+\frac{3}{2})}\left(\frac{j_{p}+2}{j_{p}+1}\right) 
           \rightarrow \prod_{p \in s} e^{-\beta^{-1} \left( j_{f}+ \frac{3}{2} \right)}(1+\mathcal{O}(\epsilon))
           \quad \text{as $j^{*} \rightarrow \infty$}
\end{eqnarray}
The notation $\mathcal{O}(\epsilon)$ denotes a quantity that is no greater than $\epsilon$ in magnitude, where $\epsilon$
is a bound\footnote{
As $\epsilon$ appears in more than one type of bound each of which requires a certain minimum $j^{*}$, we take
the maximum of the $j^{*}$ for each type of bound to control the overall limit.
}
that approaches zero as $j^{*} \rightarrow \infty$. Combining the vertex and plaquette interactions just discussed,
let us now consider a general contribution 
to $\left< O_{\Gamma_{w}} \right>_{+}^{S_{\Gamma}^{1}}$, the self-avoiding part of $\left< O_{\Gamma_{w}} \right>_{+}$:
\begin{eqnarray}
\left< O_{\Gamma_{w}} \right>_{+}^{S_{\Gamma}^{1}} &=& Z_{V}^{-1} \sum_{f \in \mathcal{J}^{+}} \sum_{s \in S_{\Gamma}^{1} } 
 D^{-1} I_{P}(f,s) I_{V}(f,s)A(f) \\ \nonumber
&=& Z_{V}^{-1}  \sum_{f \in \mathcal{J}^{+}} \sum_{s \in S_{\Gamma}^{1} } D^{-1} e^{ \beta^{-1}(\frac{3}{2}A(s) 
 + \sum_{p \in P(s)} j_{p})}(1+\mathcal{O}(\epsilon))^{|P(s)|}(1+\mathcal{O}(\epsilon))^{5|V(s)|}A(f),
\end{eqnarray}
where $P(s)$ and $V(s)$ denote the sets of plaquettes and vertices; absolute value lines around finite sets denotes
the number of elements.
To further resolve the factors that may contribute to or diminish confinement we organized the sum over worldsheets
according to their area:
\begin{eqnarray}\label{asympalaw}
\left< O_{\Gamma_{w}} \right>_{+}^{S_{\Gamma}^{1}} &=&
 Z_{V}^{-1} \sum_{f \in \mathcal{J}^{+}} \sum_{A \geq A_{\text{Min}}(\Gamma)} \sum_{s \in S_{\Gamma}^{1}(A) }  
 D^{-1} e^{ \beta^{-1}(\frac{3}{2} A(s) + \sum_{p \in P(s)} j_{p})}(1+\mathcal{O}(\epsilon))^{5|V(s)|+|P(s)|}A(f) \\ \nonumber
& \rightarrow &
 Z_{V}^{-1} \sum_{f \in \mathcal{J}^{+}} \sum_{A \geq A_{\text{Min}}(\Gamma)} \sum_{s \in S_{\Gamma}^{1}(A) }  
 D^{-1} e^{ \beta^{-1}(\sum_{p \in P(s)} j_{p})}A(f) \quad \text{as $j^{*},\beta \rightarrow \infty$.} \\ \nonumber
\end{eqnarray}
where $A_{\text{Min}}$ is the area (in number of elementary plaquettes) of a minimal surface
bounding $\Gamma$, and $S_{\Gamma}^{1}(A)$ is the restriction of the set $S_{\Gamma}^{1}$ to surfaces of
area $A$.

We have written (\ref{asympalaw}) in a way that is particularly transparent with regard to confinement in both the strong and weak coupling
limits. First, we see from the first line of~(\ref{asympalaw}) that in the $\emph{strong}$ coupling limit of
$\beta \rightarrow 0$, area law behavior will eventually dominate any roughening effects --- while the growth in the number of
surfaces of a given area is exponential in the area, it is depends only on enumerative geometry and is thus clearly independent
of $\beta$; conversely, we are free to set $\beta$ arbitrarily small and thus force a tension --- this is has long been a standard
result in the literature.

By contrast, in the $\beta \rightarrow \infty$ limit we find that the tension provided by the $e^{\beta^{-1}A}$ factor
decays to zero (for any fixed $\delta_{i}$), and thus can not provide tension to general worldsheets.  
The determining factor thus becomes the sum over $j_{p}$ over the worldsheet surface, as can be seen from the 
second line of~(\ref{asympalaw}).
Roughly speaking, if the background spin averaged over any given worldsheet grows faster than $\beta$ (which divides the sum) 
as $\beta \rightarrow \infty$, than a positive tension bounded away from zero will be present in the weak coupling limit.
This particular means of realizing area damping via the plaquette interaction is presented as a formal conjecture in Section 5.

\subsubsection{Self-intersecting worldsheets}
The set of $\Gamma$-admissible spin foams contains contributions whose resolution into background and worldsheet
require a worldsheet that intersects itself. In general, intersection can happen along vertices, edges, and plaquettes (or
any combination of these). We first consider a self-intersection that occurs at one or more vertices or edges, but not a plaquette. 
The important consideration here is that there are a limited number of ways this can happen without also changing a plaquette spin. 
Indeed, on a three-dimensional cubic lattice only one self-intersection can occur along an edge without increasing any 
plaquette spins; similarly for vertices.  Because this sort of self-intersection will generically lead to 
$\delta_{p}(s)=1$ in the vertex amplitudes of the intersection, we invoke the vertex decoupling shown above to find that these types 
of self-intersections do not provide any negative tension in the weak coupling limit.

The remaining cases consist of worldsheets in which a self-intersection occurs at a plaquette.  
For some range of self-intersection ($\delta_{i}>1$), the same arguments for the self-avoiding case can be applied.
Moreover, as $j^{*}$ gets larger this range will grow as it will take larger perturbations to appreciably change the angles
in the asymptotic formula.  However, at \emph{any} $j^{*}$ there will be some sufficiently large $\delta_{i}$ such that
the change in dihedral angles is too great to be within the assumptions of the self-avoiding case. What is required to
deal with these cases is a condition that bounds the growth of $I_{V}$ for increasing $\delta_{p}(s)$. One way this could
be true (described in more detail in Section 5) is if there existed a bound on $I_{V}$ that grew more slowly 
than the exponential decay factor that comes from $I_{P}$.  If this is the case then the cut-off $j^{*}$ could be 
raised until the $\delta_{i}$ appearing in the Type 1 factor from the $I_{P}$ were large enough to
exceed the bound on $I_{V}$, resulting in tension for these sheets. 

\subsection{Sub-asymptotic backgrounds}
Recall that for given a spin foam $f$, a sub-asymptotic plaquette is one for which $j_{p}(f) < j^{*}$, where $j^{*}$ is 
a fixed cut-off. We further define a spin foam to be sub-asymptotic if it contains one or more plaquettes that are sub-asymptotic.
Our approach in treating this case will be to find a bound for $\left< O_{\Gamma_{w}} \right>_{0}$ that is suppressed 
exponentially in the weak coupling limit.

We start by dividing the sub-asymptotic background spin foams into classes $F^{0}(N)$, each element of which is
a pair consisting of a subset of the lattice with $N$ plaquettes and an associated locally admissible 
spin foam labelling of those $N$ plaquettes by sub-asymptotic spins and edges of these plaquettes by compatible intertwiners.
We next split the sum into asymptotic and sub-asymptotic components, for both background spin foams and worldsheets, as follows:
\begin{eqnarray}\label{bigbigsum}
\left< O_{\Gamma_{w}} \right>_{0} &=& Z_{V}^{-1}  \sum_{f \in \mathcal{J}_{0}} \sum_{s \in S_{\Gamma} } D^{-1}
I_{P}(f,s) I_{V}(f,s)A(f,s) \\ \nonumber
&=& Z_{V}^{-1} \sum_{N=1}^{N=|P|}  \sum_{f^{0} \in F^{0}(N)} \sum_{s^{0}
\in S_{\Gamma}^{0}(f^{0})}  \sum_{f^{*} \in F^{*}(f^{0})}  \sum_{s^{*} 
\in S_{\Gamma}^{*}(f^{0},s^{0}) } D^{-1}(f^{0},f^{*},s^{0},s^{*})
\\ \nonumber
&\times& I_{V}(f^{0},s^{0}) I_{V}(f^{*},s^{*}) I_{P}(f^{0},s^{0}) I_{P}(f^{*},s^{*}) A(f^{0})A(f^{*}),
\nonumber
\end{eqnarray}
where the set $F^{*}(f^{0})$ consists of \emph{asymptotic completions} of the sub-asymptotic foam $f^{0}$: these are 
assignments of spins to the set of asymptotic plaquettes and edge intertwiners compatible with a given assignment of spins to 
sub-asymptotic plaquettes and edges, $f^{0}$. Given an $f^{0}$, we define another set $S^{0}_{\Gamma}(f^{0})$ that
is the restriction of the set $S_{\Gamma}$ to the sub-asymptotic plaquettes of $f^{0}$.  Finally, for any given $s^{0}$ and
$f^{0}$, the set of admissible completions of the worldsheet in asymptotic regions can be formed; we denote it 
by $S_{\Gamma}^{*}(s^{0}, f^{0})$.
Note that lower dimensional simplices at boundaries (edges and vertices with both asymptotic and sub-asymptotic 
incident plaquettes) between asymptotic and sub-asymptotic regions are defined to be in the sub-asymptotic
component.

A considerable obstacle to forming a useful bound on $\left< O_{\Gamma_{w}} \right>_{0}$ is the fact
that $A(f)$ can take on negative amplitudes. In particular, in forming upper bounds we will be careful not 
to take absolute values of amplitudes being summed over asymptotic regions of the state space, as quantities of this
type can grow exponentially in the volume of the lattice faster than $Z_{V}$ itself (this manifests
itself in numerical simulation as the decay of the sign expectation value), where cancellation
occurs between positive and negative amplitudes.
In contrast, for sub-asymptotic quantities at a fixed cut-off, we shall see that there arises factors that are
bounded by exponential factors in the volume at a rate that is $\beta$ independent. We will show how they can be suppressed by
terms coming from asymptotic factors in the $\beta \rightarrow \infty$ limit. In summary, in proceeding we will
take absolute values on sub-asymptotic quantities but not with asymptotic quantities.

We will start by considering the sub-asymptotic interactions and amplitude that appears 
in~(\ref{bigbigsum}): $I_{V}(f^{0},s^{0})$, $I_{P}(f^{0},s^{0})$,
and $A(f^{0},s^{0})$.  Recall that $I_{V}(f^{0},s^{0})$ can be written as a product of five factors of the form
$\frac{  \{ \vec{j}_{v}(f) + \vec{\delta}(s) \} }{ \{ \vec{j}_{v}(f) \} }$, a ratio of a $6j$ symbols\footnote{
The statements made of $6j$ symbols in this section (essentially boundedness on finite sets of arguments) are meant 
to include the modified $6j$ symbols at $\Gamma$ as well as the original unmodified $6j$.
}
with background
spins displaced by the worldsheet and the original un-displaced $6j$ symbol.
In the case where the $6j$ symbol contains all arguments less than $j^{*}$, than the interaction associated with the 
ratio $\frac{  \{ \vec{j}_{v}(f) + \vec{\delta}(s) \} }{ \{ \vec{j}_{v}(f) \} }$ is bounded by
\begin{eqnarray}\label{bounded1}
\mathcal{M}_{1} \equiv |( \text{min} \left[ \{ \vec{j} \} \right]_{j<j^{*}} )^{-1}|,
\end{eqnarray}
the inverse of the smallest value taken by the $6j$ symbol on the finite set of values 
satisfying $j < j^{*}$ for all arguments $j$ (because the numerator is a $6j$ symbol it is simply bounded by unity --- see
Appendix A).

In the case where $6j$ symbols contain a mixture of asymptotic and sub-asymptotic arguments, we assume (currently
based on numerical experiments rather than rigorous proof) that there is a finite range of values $J_{0}$ of the 
asymptotic arguments, beyond which the $6j$ symbol becomes monotonically decreasing, and hence gives a 
ratio of $\frac{  \{ \vec{j}_{v}(f) + \vec{\delta}(s) \} }{ \{ \vec{j}_{v}(f) \} }<1$.
In which case we have
\begin{eqnarray}\label{bounded2}
\mathcal{M}_{2} \equiv |( \text{min} \left[ \{ \vec{j} \} \right]_{j<j^{*},J_{0}} )^{-1}|.
\end{eqnarray}
where the minimization is over all sub-asymptotic subsets of arguments, and for each sub-asymptotic subset, a minimum
over the bounded non-monotonic range of the remaining arguments $J_{0}$. 
Defining $\mathcal{M} = \text{max}[\mathcal{M}_{1},\mathcal{M}_{2}]$, we thus have
\begin{eqnarray}
|I_{V}(f^{0},s^{0})| \leq \mathcal{M}^{5v_{0}(s^{0})},
\end{eqnarray}
where the function $v_{0}(s^{0})$ is the number of vertices with some incident plaquettes labelled by sub-asymptotic spins; 
$\mathcal{M}$ is raised to the fifth power as there are that many $6j$ symbols in the vertex amplitude.
We shall also make use of the following inequalities for the remaining sub-asymptotic factors:
\begin{eqnarray}
|A(f^{0},s^{0})| \leq P_{1}, \quad |I_{P}(f^{0},s^{0})| \leq P_{2},
\end{eqnarray}
which follows from the fact that $6j$ symbols are bounded by unity; as well, the plaquette amplitude by inspection
has a maxima and is bounded from below in the region where it is not monotonically decreasing with displacements by $\delta_{i}$.
The maxima always exists because the exponential parts of $I_{P}$ will eventually damp out the linear increasing dimension factor for any
non-zero $\beta$. Note that the $P_{1}$ and $P_{2}$ are $f^{0}$ and $s^{0}$ dependent but we suppress this for compactness in notation.

We turn now to the asymptotic summations and factors. For a given choice of $f^{0}$ and $s^{0}$, we consider
the factor contributed by summing over all completions of the asymptotic part:
\begin{eqnarray}
\mathcal{T}(f^{0},s^{0}) \equiv
\sum_{f^{*} \in F^{*}(f^{0})}  \sum_{s^{*} 
\in S_{\Gamma}^{*}(f^{0},s^{0}) } D^{-1}(f^{0},f^{*},s^{0},s^{*})I_{V}(f^{*},s^{*})I_{P}(f^{*},s^{*})A(f^{*}).
\end{eqnarray}
We shall refer to $\mathcal{T}(f^{0},s^{0})$ as the \emph{asymptotic tail}. Using this definition
and introducing the sub-asymptotic bounds above, we can substitute into~(\ref{bigbigsum}) and thus bound $\left< O_{\Gamma_{w}} \right>_{0}$
as follows:
\begin{eqnarray}\label{sub_asy_bound}
\left< O_{\Gamma_{w}} \right>_{0} &\leq&  Z_{V}^{-1} \sum_{N=1}^{N=|P|}  \sum_{f^{0} \in F^{0}(N)} \sum_{s^{0}
\in S_{\Gamma}^{0}(f^{0})} \mathcal{M}^{5v_{0}(s^{0})}(P_{1}P_{2})^{p_{0}(s)} \mathcal{T}(f^{0},s^{0}).
\end{eqnarray}
We next exhibit $\mathcal{T}(f^{0},s^{0})$ in terms of an average over a \emph{restricted} partition function,
which will suggests a conjecture that is critical for sub-asymptotic suppression in the weak coupling limit. We define
\begin{eqnarray}\label{restricted}
Z^{*}(f^{0}) = \sum_{f^{*} \in F^{*}(f^{0})}A(f^{*}),
\end{eqnarray}
which can be thought of as a \emph{restricted} partition function that sums over all foams which have a
fixed sub-asymptotic component $f^{0}$. We now write
\begin{eqnarray}\label{a_wsheet_weight}
\mathcal{T}(f^{0},s^{0}) &=& Z^{*}(f^{0})\frac{
\sum_{f^{*} \in F^{*}(f^{0})}  \sum_{s^{*} 
\in S_{\Gamma}^{*}(f^{0},s^{0}) } D^{-1}(f^{0},f^{*},s^{0},s^{*})I_{V}(f^{*},s^{*})I_{P}(f^{*},s^{*})A(f^{*})
}{ \sum_{f^{*} \in F^{*}(f^{0})}A(f^{*})  } \\ \nonumber
&=& Z^{*}(f^{0}) \sum_{A=\alpha(f^{0},s^{0})}^{\infty}\mathcal{O}(e^{-\tau A}),
\end{eqnarray}
where $\alpha$ is the smallest area of the restrictions of a $\Gamma$-filling surfaces to the asymptotic region; 
in general it depends on the position of $\Gamma$ relative to the asymptotic plaquettes (e.g. $A_{\text{Min}}$
if $\Gamma$ is contained entirely in a region of asymptotic plaquettes) and to how the sub-asymptotic 
worldsheet $s^{0}$ ends on asymptotic regions. The sum up to ``$\infty$'' denotes that self-intersecting
worldsheets (of unbounded amount of total spin) are also included. In writing~(\ref{a_wsheet_weight}) we 
have assumed a generalized form of asymptotic confinement: summing over all completions on (a possibly multiply 
connected set of asymptotic plaquettes) can be organized by area class, each with its own (positive) tension that
is bounded away from zero by some ($\tau^{*}$).
Given this assumption, the sum converges, as it is a sum of terms bounded by exponentials with negative arguments 
proportional to area $A$.

Inserting~(\ref{a_wsheet_weight}) into~(\ref{sub_asy_bound}) we have
\begin{eqnarray}
\left< O_{\Gamma_{w}} \right>_{0} &\leq&   \sum_{N=1}^{N=|P|}  \sum_{f^{0} \in F^{0}(N)} \sum_{s^{0} \in S_{\Gamma}^{0}(f^{0})} 
\mathcal{M}^{5v_{0}(s^{0})}(P_{1}P_{2})^{p_{0}(s)} \left( \frac{|Z^{*}(f^{0})|}{Z_{V}} \right)
\sum_{A=\alpha(f^{0},s^{0})}^{\infty}\mathcal{O}(e^{-\tau A}).
\end{eqnarray}
Let any $N$ and Wilson loop $\Gamma$ be given. For foams of a given fixed sub-asymptotic component $f^{0} \in F^{0}(N)$, 
let us characterize the possible terms in the in the inner sum:
\begin{eqnarray}
\sum_{s^{0} \in S_{\Gamma}^{0}(f^{0})} 
\mathcal{M}^{5v_{0}(s^{0})}(P_{1}P_{2})^{p_{0}(s)}\left( \frac{Z^{*}(f^{0})}{Z_{V}} \right)
\sum_{A=\alpha(f^{0},s^{0})}^{\infty}\mathcal{O}(e^{-\tau A}).
\end{eqnarray}
We recall that the minimal total area (both asymptotic and sub-asymptotic) of a worldsheet must be at least
the area of the minimal surface spanning $\Gamma$.
We see by inspection that the sub-asymptotic worldsheet interaction factor is bounded by the number
of vertices (and hence plaquettes) in the asymptotic region --- that is, there is a factor of negative tension 
\begin{eqnarray}
\mathcal{M}^{5v_{0}(s^{0})}(P_{1}P_{2})^{p_{0}(s)}=e^{p_{0}(s)\log(P_{1}P_{2})}e^{\log(\mathcal{M})5v_{0}(s^{0})} 
\leq e^{T(j^{*})A_{0}(s^{0})}
\end{eqnarray}
where $A_{0}(s^{0})$ is the number of plaquettes or vertices (whichever is greater)
occupied by the sub-asymptotic component of the worldsheet $s^{0}$ and $T(j^{*}) \equiv \log(P_{1}P_{2})+5\log(\mathcal{M})$.
This bounding factor certainly does not provide confinement as it has precisely the opposite effect: the weight of 
worldsheets $s^{0}$ \emph{increases} exponentially with its area. However, the factor $\frac{Z^{*}(f^{0})}{Z_{V}}$ 
plays a crucial countervailing role. We consider the conjecture that the growth rate of the restricted partition 
function is slower by a factor proportional to the number of sub-asymptotic 
plaquettes (these are essentially degrees of freedom that are frozen) $A_{0}(f_{0})$, that is:
\begin{eqnarray}\label{freeze_ineq}
\frac{|Z^{*}(f^{0})|}{Z_{V}} < e^{-\mu \beta A_{0}(f^{0})} \leq e^{-\mu \beta A_{0}(s^{0})}
\end{eqnarray}
based on the assumption that both restricted and unrestricted partition functions increase exponentially in $\beta$ and in
plaquette volume. Because $A_{0}(s^{0}) \leq A_{0}(f^{0})$, we can form a bound that is directly in terms of $s^{0}$ and we do so on the
rightmost side of~(\ref{freeze_ineq}). We then apply this inequality to find:
\begin{eqnarray}
\left< O_{\Gamma_{w}} \right>_{0} &\leq&   \sum_{N=1}^{N=|P|}  \sum_{f^{0} \in F^{0}(N)} e^{-\mu \beta A_{0}(f^{0})} 
\sum_{s^{0} \in S_{\Gamma}^{0}(f^{0})} e^{T(j^{*})A_{0}(s^{0})} \sum_{A=\alpha(f^{0},s^{0})}^{\infty}\mathcal{O}(e^{-\tau A}) 
\\ \nonumber
 &=&  \sum_{N=1}^{N=|P|}  \sum_{f^{0} \in F^{0}(N)} e^{-\frac{\mu}{2} \beta A_{0}(f^{0})} 
\left( \sum_{s^{0} \in S_{\Gamma}^{0}(f^{0})} 
e^{(T(j^{*})-\frac{\mu}{2}\beta)A_{0}(s^{0})} \sum_{A=\alpha(f^{0},s^{0})}^{\infty}\mathcal{O}(e^{-\tau A}) \right).
\end{eqnarray}
Consider next the quantity parenthesized above, which we shall define as $\mathcal{S}(f^{0})$:
\begin{eqnarray}
 \mathcal{S}(f^{0}) &\equiv& \sum_{s^{0} \in S_{\Gamma}^{0}(f^{0})} 
e^{(T(j^{*})-\frac{\mu}{2}\beta)A_{0}(s^{0})} \sum_{A=\alpha(f^{0},s^{0})}^{\infty}\mathcal{O}(e^{-\tau A}) \\ \nonumber 
&=& \sum_{A \in A_{\text{Min}}(f^{0})}^{\infty}\sum_{s^{0} \in S_{\Gamma}^{0}(A)} 
e^{(T(j^{*})-\frac{\mu}{2}\beta)A_{0}(s^{0})} \sum_{A=\alpha(f^{0},s^{0})}^{\infty}\mathcal{O}(e^{-\tau A}) \\ \nonumber
& \leq_{\beta \rightarrow \infty} &
\sum_{A=A_{\text{Min}}}^{\infty}\mathcal{O}(e^{-\tau A}) 
\end{eqnarray}
with the last inequality holding for sufficiently large $\beta$; the inequality is saturated when $\beta$ is 
tuned so as to provide tension $\tau$ equivalent to the asymptotic tension and becomes a strict upper bound
for larger $\beta$. Note this last bound is independent of $f^{0}$.

We shall now introduce a number of bounds to control the growth of the remaining summation over $N$ and $F^{0}(N)$.
First, the number of partitions of
plaquettes into $N$ sub-asymptotic and $|P|-N$ asymptotic groupings is simply $\binom{|P|}{N}$. We also observe 
that for a given number $N$ of sub-asymptotic plaquettes, there are no more than $(j^{*})^N$ labellings of the
sub-asymptotic plaquettes (of course not all will be components of admissible spin foams).

We now combine the various bounds introduced above to find an overall bound on $\left< O_{\Gamma_{w}} \right>_{0}$.
Recalling that $N = A_{0}(f^{0})$ is the number of sub-asymptotic plaquettes, we have
\begin{eqnarray}\label{eqn:suppression}
\left< O_{\Gamma_{w}} \right>_{0} &\leq_{\beta \rightarrow \infty}&   \sum_{N=1}^{N=|P|}  \sum_{f^{0} \in F^{0}(N)} e^{-\frac{\mu}{2}
\beta A_{0}(f^{0})} \sum_{A=A_{\text{Min}}}^{\infty}\mathcal{O}(e^{-\tau A}) \\ \nonumber
&\leq& \sum_{N=1}^{N=|P|} \binom{|P|}{N}(j^{*})^{N} e^{-\frac{\mu}{2} \beta N} 
\sum_{A=A_{\text{Min}}}^{\infty}\mathcal{O}(e^{-\tau A}) \\ \nonumber
&\leq& \sum_{N=1}^{|P|} e^{ (\log(|P|)+\log(j^{*}))N } e^{-\frac{\mu}{2} \beta N} \sum_{A=A_{\text{Min}}}^{\infty}\mathcal{O}(e^{-\tau A})
\end{eqnarray}
where we've used the inequality $\binom{|P|}{N} < |P|^N = e^{\log(|P|)N}$.
We assume that in the limit of large $\beta$ invoked here, the lattice volumes (as measured by number of plaquettes $|P|$)
that fall within the scaling window grow no faster than exponential in $\beta$, and hence the theory has the appropriate 
critical behavior to provide a continuum limit. 

Finally, we state our conclusion for the sub-asymptotic case: subject to the hypothesis given in~(\ref{freeze_ineq}) for sufficiently large $\beta$, sub-asymptotic
contributions 
\begin{enumerate}
\item{Are \emph{bounded} by a decaying exponential series with leading order having tension associated
with asymptotic tension. This series converges.}
\item{The coefficient in front of the convergent series over worldsheet areas is a decaying exponential for sufficiently 
large $\beta$, for any lattice volume (as measured by number of plaquettes). The rate of decay is also proportional to 
the number of sub-asymptotic plaquettes $N$.}
\end{enumerate}
On account of point (2), we conclude that in the weak coupling limit, \emph{all sub-asymptotic contributions are suppressed}:
$\left< O_{\Gamma_{w}} \right>_{0}$ at a fixed cutoff $j^{*}$ (e.g. as given by some bound $\epsilon$) can be made 
arbitrarily small for sufficiently large $\beta$.

\section{Conjectures}
In Section 4 we described scenarios in which area law behavior for asymptotic backgrounds could emerge, and how sub-asymptotic
backgrounds are suppressed --- both in the weak coupling limit. In this section we collect together more precise
statements of those properties that are sufficient to provide area law behavior, as well as an additional
\emph{balance condition}, which if true would overcome the technical problems raised by the non-positivity 
of the spin foam amplitudes. 

Below we have presented these conjectures according to the general structure proposed in~\cite{Conf1}.  That is,
we start by conjecturing that the growth rate of background spins is of the same or high order as that
of $\beta$ itself in the $\beta \rightarrow \infty$ limit.  We then introduce sub-asymptotic suppression to argue 
that asymptotic background foams are the only ones relevant in the weak coupling limit. Next we take up the question of 
vertex stability on asymptotic backgrounds, which breaks up into two cases --- self-avoiding and self-intersecting worldsheets.  
In the self-avoiding case, we conjecture that the Ponzano-Regge asymptotics may
be true in a uniform sense, which would (by the analysis of Section 3) be sufficient to establish vertex stability in 
the self-avoiding section (for all asymptotic spin foams).  To treat the remaining self-intersecting sheets,
we conjecture the existence of a lower bound on the $6j$ symbol (outside regions of monotonic decrease in $\delta_{p}$)
that is sub-exponential in $\delta_{p}$ (introduced in Section 3). Such a bound would be sufficient to provide vertex 
stability, as the relevant part of the plaquette interaction decays exponentially in $\delta_{p}$.

Given the preceding conjectures, our final conjecture is that the balance condition introduced 
in~\cite{Conf1} is true, allowing us to form an area-law upper found on the growth the Wilson loop expectation value.

\subsection{Area damping from plaquette interaction}\label{plaq_conj}
Considering both avoiding and self-avoiding sheets, we define the plaquette tension as follows:
\begin{eqnarray}\label{plaq_tension}
I_{P}(f,s) &=& \left(e^{\beta^{-1} \sum_{p \in s} \delta_{p}(s) j_{p} } \right)
\left( e^{-(2\beta)^{-1}\sum_{p \in s} \delta_{p}(s)(\delta_{p}(s)+2)} \right)
\left(\prod_{p \in s} \frac{j_{p}+1+\delta_{p}(s)}{j_{p}+1} \right)  \\ \nonumber
           &\equiv& e^{-\tau_{p}(f,s)A(s)},
\end{eqnarray}
and apply the definition of area damping from plaquette interaction introduced in~\cite{Conf1}
\begin{equation}\label{damping_cdtn}
\frac{\sum_{f \in F_{V}} \left[ \sum_{A \geq A_{\text{Min}}(\Gamma)} \sum_{s \in S_{\Gamma}(A)} D^{-1}(f,s)e^{-\tau_{p}(f,s)A}
\right] A(f) }
{\sum_{f \in F_{V}} A(f)} < \infty \quad \text{as $\beta \rightarrow \infty$}
\end{equation}
Define the first, second, and third parenthesized factors of~(\ref{plaq_tension}) to be Type 1, 2, and 3 factors,
respectively.
At any fixed $\beta$ and background foam $f$, the Type 2 factor will eventually damp out growth in the 
Type 3 factor. However, as $\beta \rightarrow \infty$, the number of terms in the Type 3 factor that
could give above unity overall $I_{P}$ grows without bound due to the weakening of the Type 2 factor
with inverse $\beta$.

By elimination, we conjecture that the Type 1 factor is critical to obtaining an area law. We observe that on backgrounds in 
which the average spins over any sheet is greater than $\beta$, the first factor
will provide a positive tension to all worldsheets, with self-intersecting ones being suppressed at a rate proportional
to their $\delta(s)$.  As the asymptotic cut-off scale is increased, the slope of the linear growth in $\delta$ in
the Type 3 factor will insufficient to overcome the tension from Factor 1.
Thus for sufficiently large cut-off it is possible for $I_{P}$ to provide a sub-unity 
(positive tension factor) in the $\beta \rightarrow \infty$ limit for a large class of background
spin foams.  Thus, as long as backgrounds of this type tend to dominate contributions to the expectation value
in the $\beta \rightarrow \infty$ limit, the area damping condition will be met.  More succinctly:
\begin{conj}
As $\beta \rightarrow \infty$, the set of background spin foams for which the Type 1 factor provides a
positive tension sufficient to satisfy~(\ref{damping_cdtn}) dominates the contributions to $\left<O_{\Gamma_{w}}\right>$.
\end{conj}
   This is simply a more specific form of the general area damping condition given in~\cite{Conf1}, as it identifies
the means by which the defining inequality~(\ref{damping_cdtn}) is realized.
   We remark in passing that this conjecture is well-suited to numerical simulation --- the expectation value of the Type 1 factor
on the vacuum ensemble can be computed using both conventional and dual spin foam algorithms~(as in~\cite{CCK}) and is the
topic of ongoing investigation by the author. 

\subsection{Sub-asymptotic suppression}\label{suppress_conj} Let $Z^{*}(f^{0})$ be defined as in equation~(\ref{restricted}). We
conjecture that 
\begin{conj}
For any sub-asymptotic spin foam component $f^{0}$ we have
\begin{eqnarray}
\frac{|Z^{*}(f^{0})|}{Z_{V}} < e^{-\mu \beta A_{0}(f^{0})}
\end{eqnarray}
where $A_{0}(f^{0})$ is the number of plaquettes in $f^{0}$ and $\mu$ is a positive constant.
\end{conj}
The number of degrees of freedom summed over in forming $Z_{V}$ and its restrictions $Z^{*}(f^{0})$ is proportional to the volume of the
complement of $f^{0}$.

Testing this conjecture is possible by dual spin foam methods~\cite{CCK} by studying the characteristic observables 
of $f^{0}$ --- functions that are unity on background foams $f$ where spins are consistent with $f_{0}$ and zero otherwise.
It would also be interesting to investigate analytic or numerical methods of showing the conjecture that used 
the integral presentation of the partition function as is done in conventional lattice gauge theory.

\subsection{Vertex stability: self-avoiding worldsheets}
The following conjecture allows us to use an asymptotic form of the vertex amplitude for large
values of spin $j$.
\begin{conj}[Asymptotic formula with uniform cut-off] \label{asy_uni_conj} Let $\epsilon > 0$ be given. Then there exists
some $j^{*}$ such that if $j_{i} > j^{*}$ for all of the $j_{1} \ldots j_{6}$ arguments to the $6j$ symbol
$\{ \vec{j}_{v} \}$ then
\begin{eqnarray}
\frac{\{ \vec{j}_{v} \}}{\text{\emph{Asy}}(\vec{j}_{v})} = 1 + \epsilon(j^{*})
\end{eqnarray}
where
\begin{eqnarray}\label{asy_vertex}
\text{\emph{Asy}}(\vec{j}_{v}) \equiv \begin{cases} \sqrt{\frac{2}{3\pi V}}\cos\left\{ \sum(j_{i}(v)+1)
\frac{\theta_{i}}{2}+\frac{\pi}{4}\right\} \quad \text{for Euclidean tetrahedra.} \\ 
\text{Otherwise, decaying exponentially.}\end{cases}
\end{eqnarray}
where the Euclidean condition and other aspects of the asymptotic formula are discussed in Section 3.
\end{conj}
This allows one to analyze the asymptotic case of self-avoiding worldsheets as in Section 4.2.1,
culminating in~(\ref{asympalaw}), at which point vertex stability (for that class of worldsheet)
reduces to the area damping conjecture.

As mentioned above (Section~\ref{sec:largespin}), rigorous derivations of the 
asymptotic formula in the literature are presently limited to rescaling limits, where a set of admissible
arguments is rescaled. Recent advances in analyzing $6j$ symbols~\cite{GVdW09,AA09} and in particular ~\cite{LJ09}
are encouraging --- we expect it may be known soon rather the Ponzano-Regge formula or some generalization thereof
can provide a uniform bound that behaves in the way stated.

We should note here the significance of the vertex interaction going to unity (rather than above or below unity,
which would provide negative and positive tension factors). One implication is that for self-avoiding worldsheets
on asymptotic backgrounds, confinement must by elimination come from the plaquette interaction. Further, assuming
confinement is true, this result can be used as a starting point (base case) in showing that the self-intersecting
case also has decoupled or at least provides bounded negative tension. We turn to this case next.

\subsection{Vertex stability: self-intersecting case}
In~\cite{Conf1} a relatively ``high-level'' definition of vertex stability is given, with the idea that different types
of estimates may be needed for models of different gauge groups and dimensionality. An area averaged tension 
is defined as
\begin{equation}
e^{-\bar{\tau}(f,A)A} \equiv \frac{\sum_{s \in S_{\Gamma}(A)}I_{V}(f,s)I_{P}(f,s)}{e^{T_{0}(A-A_{\text{Min}}) }} 
\end{equation}
this allows comparison of the interaction factor with the ``negative tension'' arising from the
exponential growth in surface number, $n(A) = e^{T_{0}(A-A_{\text{Min}})}$, leading to the condition
that 
\begin{equation}
\left< O_{\Gamma_{w}} \right> = \frac{\sum_{f \in F_{V}} 
\left[ e^{-T_{0}A_{\text{Min}}} \sum_{A \geq A_{\text{Min}}}
e^{ \left( T_{0}-\bar{\tau}(f,A) \right) A} \right] A(f)}
{\sum_{f \in F_{V}}A(f)},
\end{equation}
indicating that
\begin{equation}
\bar{\tau}(f,A)-T_{0} > 0
\end{equation}
is the characteristic inequality. Showing that this inequality is satisfied such that the infinum
of $\bar{\tau}(f,A)$ over all $f$ and $A$ that aren't suppressed as $\beta \rightarrow \infty$ then defines
the condition of vertex stability for a model.

A general approach to establishing the above is to carry out a local analysis, in the sense that
if $|I_{V}(f,s)|$ is above unity over a worldsheet, then some factor of which it is a product must be over unity.
Thus, if one considers all admissible arguments to a factor of $|I_{V}(f,s)|$ at a \emph{single} vertex and
establishes bounds on that quantity for all admissible arguments, one could extend it to a bound over 
any worldsheet for any asymptotic background.

In our present case we can pursue this local strategy, and moreover exploit the fact that vertex interaction factors are directly
related to perturbations of $6j$ symbols. The availability of asymptotic estimates then provides a means of
realizing the necessary condition, leading us to conjecture
\begin{conj}
We define \emph{sub-exponential in $\delta_{i}$} to mean that $\frac{|I_{V}|(f,s)}{exp(T\sum_{i}\delta_{i}))} \rightarrow 0$
for any $T > 0$; the $\delta_{i}$ can be any collection of the $\delta_{i}$ arguments appearing in~(\ref{6j_args}).
For all possible admissible arguments where $I_{V}$ is non-zero, one of the two is true:
\begin{enumerate}
\item{Monotonic decreasing with $\delta_{i}$ or}
\item{$|I_{V}(f,s)|$ is sub-exponential in $\delta_{i}$.}   
\end{enumerate}
\end{conj}
The expectation behind this conjecture is that in the Euclidean regime where the $6j$ asymptotic
is characteristically oscillatory (multiplied an effective power law volume factor) there will 
exist some sub-exponential bound away from zero. This will in turn limit the rate at which
displacements of $\delta_{i}$ can lead to growth relative to undisplaced $6j$ symbol; if so, then
the exponential damping from the Type 1 plaquette factor will be able to attenuate worldsheets
with arbitrarily high levels of self-intersection.

In the non-Euclidean cases where the $6j$ may  exhibit exponential decay in the sum over $j_{i}\theta_{i}$, large displacements of
the $6j$ factors in $|I_{V}|$ will be less than unity if $6j$ is monotonic decreasing.
Monotonic decay when increasing single arguments or some subset of the arguments has (to the author's knowledge)
not yet been rigorously shown, but is well-supported by numerical testing by the author and in 
the literature.

\subsection{Balance relations among positive and negative sign contributions}
As emphasized in~\cite{Conf1}, the amplitude $A(f)$ is in general non-negative for dual spin foam models of Yang-Mills
theories and that is certainly true of the present case; this raises some considerable technical complications
in forming an upper bound.
We next make the following conjecture: 
\begin{conj}
\emph{The interaction factor for any given worldsheet $s$, averaged over negative
amplitude states $F_{V}^{-}$ approaches its average over positive amplitude states $F_{V}^{+}$ as $\beta \rightarrow \infty$}.
\end{conj}
What this allows us to do is take either the positive or negative weight background spin foams to estimate
the Wilson loop expectation value. Taking the positive weight configurations:
\begin{eqnarray}\label{final_bound}
\left< O_{\Gamma_{w}} \right>
&\rightarrow&
\frac{\sum_{f \in F_{V}^{+}}
e^{-T_{0}A_{\text{Min}}}
\sum_{A \geq A_{\text{Min}}}
e^{ \left( T_{0}-\bar{\tau}(f,A) \right) A}
|A(f)|}{\sum_{f \in F_{V}^{+}}|A(f)|} \quad \text{as $\beta \rightarrow \infty$}
\\ \nonumber
&<& 
\frac{
\left(
e^{-T_{0}A_{\text{Min}}}
\sum_{A \geq A_{\text{Min}}}e^{-\tau^{*}A}
\right)
\sum_{f \in F_{V}^{+,\tau}} |A(f)|} {\sum_{f \in F_{V}^{\tau^{*}+}}|A(f)|} + E(\beta) \\ \nonumber
&=& e^{-T_{0}A_{\text{Min}}} \sum_{A \geq A_{\text{Min}}}e^{-\tau^{*}A} + E(\beta)
\\ \nonumber
\end{eqnarray}
where the bounding tension is defined as $\tau^{*} = \text{inf}_{F_{V}^{\tau^{*}+},A} \left[ \bar{\tau}(f,A)-T_{0} \right]$.
The term $E(\beta)$ represents contributions where the existence of a bounding tension does not hold, but which can
be shown to go to zero in the $\beta \rightarrow \infty$ limit.

An alternative to the balance condition is to conjecture that there is a subset
of $\mathcal{F}_{\text{sp}}$ of stationary phase which dominates the partition function in 
the continuum limit. In this case, assuming numerator of the expectation value (as a sum
of $A(f)$ weighted by the worldsheet observable) is also dominated by configurations from 
$\mathcal{F}_{\text{sp}}$, then we can use positivity on the set $\mathcal{F}_{\text{sp}}$ to 
form a similar bound. The availability of asymptotic estimates for $A_{V}$
combined with the suppression of sub-asymptotic configurations may render a stationary phase tractable,
and is the subject of ongoing work by the author.

\subsection{Final steps}
The conjecture~(\ref{final_bound}) provides a presentation of the expectation value for
a single fixed Wilson loop $\Gamma$ as a decaying exponential sum over classes of worldsheets
of increasing area.  
To consider confinement, one must consider the behavior across a range of Wilson loops corresponding
to increasing spatial separation (and thus increased area) of quark-antiquark pairs.
As discussed in~\cite{Conf1}, one way to provide this is to define
\begin{equation}
\hat{\tau}_{\mathcal{G}} \equiv \text{inf}_{\mathcal{G}}[\tau^{*}(\Gamma)].
\end{equation}
The (leading order) of each world-sheet area expansion of $\left< O_{\Gamma_{w}} \right>$ 
can thus be bounded by a function proportional to $e^{-\tau^{*}A_{\text{Min}}}$, thus yielding the
area-law decay characteristic of a confining theory.
It should be noted that such a bound does not (by itself) rule out local variations in the expectation value
that may fluctuate within the exponentially decaying envelope.

We mention in passing how a similar analysis could be applied to the mass gap observable (2-point correlation function),
and in principle higher order correlation functions. As discussed in Appendix B of~\cite{Conf1}, the 
main point of departure is that the topology of worldsheets is different --- the background foams are unchanged
however the worldsheets terminate on two fundamental plaquettes rather than single loop $\Gamma$. As 
a result, the rate of growth with increasing area may expected to differ somewhat, as would the effective
tension of a given area class.

\section{Conclusions}
As described in~\cite{Conf1}, a Wilson loop expectation value in conventional lattice gauge theory can be
expressed as the expectation value of a certain type of observable on the vacuum spin foam state sum. This observable
is a sum of interaction factors --- one for each worldsheet bounding $\Gamma$ (modulo a degeneracy factor).
Each interaction factor is manifestly a product of factors over plaquettes and vertices of the worldsheet.
Due to the factorization of amplitudes over worldsheets and the ability to organize worldsheets by increasing area,
this spin foam formalism is particularly well-suited for identifying the absence or presence of area law decay.
In the present paper, we introduced the interaction factors associated to the 
choices $G=SU(2)$ and space-time dimension $D=3$ specifically and described methods for understanding their behavior
in different sectors: sub-asymptotic, asymptotic with self-intersecting worldsheets, and asymptotic with self-avoiding
worldsheets.

The resulting analysis shares a number of similarities with the study of lattice gauge theory through strong coupling diagrams
in the strong-coupling limit. The weak coupling case considered here however requires \emph{a priori} consideration
of all vacuum spin foam backgrounds as well as all worldsheets bounding $\Gamma$, including those with arbitrary amounts
of self-intersection. By contrast, the theory at strong coupling depends on a much more restricted class of diagrams.

The scenario analyzed here follows the general ansatz described in~\cite{Conf1}, where background foams
that dominate the weak coupling limit provide tension to worldsheets through plaquette interaction factors.
In order for this scenario to hold true, the \emph{stability of vertex interaction} property needs to be shown.
A remarkable result of the present work is that vertex stability for \emph{self-avoiding} worldsheets 
follows immediately from a well-known asymptotic formula for $6j$ symbols, of which the vertex interaction is a ratio of products.
For worldsheets of arbitrary self-intersection, we explain how the vertex interaction factor (and thus,
the $6j$ symbols that comprise it) can be bounded in such a way that the plaquette interaction can still provide
a tension.

We further show that subject to a conjecture on the relative growth rate of restricted
sums relative to the full partition function, backgrounds with sub-asymptotic spins (below a given cut-off) can
be ignored in the weak coupling limit.  This allows the remaining conjectures to be described in terms of the
asymptotic form of the vertex amplitudes, which should be susceptible to a variety of analytic methods.

Despite the above results, the picture currently presented by this approach is incomplete in several
respects. The behavior of local amplitudes critical for vertex stability involves several properties relating to the
asymptotic behavior of $6j$ symbols which, while suggested anecdotally by numerical experiment, remains to
be proven rigorously for all admissible values.
At the level of the statistics of the model, the growth rate of background spins relative to $\beta$ remains
to be shown. Further statistical results are also require to assess the growth rate of restricted relative
to unrestricted partition functions, the essential ingredient of sub-asymptotic suppression.
Finally, the presence of alternating signs continues to cloud the picture, and it will be interesting to see
if stationary phase methods may shed light on this difficulty.

In summary, the present work gives a working example of how spin foams can be used to arrive at a gauge-invariant 
understanding of confinement at weak coupling similar in concept to the standard proof of confinement at strong coupling.
To provide a concrete scenario for confinement, a number of specific properties of the spin foam model with varying
levels of heuristic and numerical support were conjectured which taken together are sufficient for an area
law bound. We hasten to add that there could be a number of alternative realizations should one of these conjectures
fail --- nonetheless, we expect that the general framework we have illustrated
will increasingly constrain the behavior local amplitudes and statistical behavior of the model as more results
(both negative and positive) are obtained, to the point where alternative possibilities can be systematically explored.  
We do hope the results obtained thus far are sufficient to spark further progress on the conjectures given here as
well as motivate similar investigations in four dimensions and for $G=SU(3)$.

\begin{acknowledgement*}
The author would like to thank Dan Christensen, Florian Conrady, and Igor Khavkine for valuable discussions that
influenced this work.
\end{acknowledgement*}

\begin{appendices}

\section{Absolute Convergence of vacuum and $\Gamma$-admissible spin foam state-sum}
\begin{thm}
\textbf{(Absolute convergence of $Z$).} The partition function $Z_{||}$, in which every the absolute value of 
every summand in the original partition function is summed, exists for any (finite) lattice $\kappa$. 
\end{thm}
The Racah-Wigner $6j$ symbols used in the present work are bounded by unity, shown in~\cite{AA09,FL02} for example.
Therefore we can write:
\begin{equation}
Z_{||}(\beta,\kappa)\leq\sum_{f\in F_{\kappa}}\left(\prod_{i\in U_{f}}\left|\left\{ \begin{array}{ccc}
j_{1} & j_{2} & j_{3}\\
j_{4} & j_{5} & j_{6}\end{array}\right\} \right|(2j_{i}+1)e^{-\frac{2}{\beta}j_{i}(j_{i}+1)A_{i}}\right)
\leq\sum_{f\in F_{\kappa}}\left(\prod_{i\in U_{f}}(2j_{i}+1)e^{-\frac{2}{\beta}j_{i}(j_{i}+1)A_{i}}\right).
\end{equation}
In the final step we replace a sum of products by a product of sums.
Each sum in the product can in turn be bounded by an integral, whose
existence is guaranteed by the finiteness of integrating a polynomial
against a gaussian on $\mathbb{R}^{+}:$
\begin{equation}
Z_{||}(\beta,\kappa)\leq\prod_{f\in\kappa}\left(\sum_{j_{f}^{'}=0}^{\infty}(j_{f}^{'}+1)e^{-\frac{1}{\beta2}j_{f}'({j'}_{f}^{'}+2)A_{i}}\right)
\leq\left(\int_{0}^{\infty}dj_{f}^{'}(j_{f}'+1)e^{-\frac{1}{2\beta}j_{f}'(j_{f}^{'}+2)}\right)^{\left|F\right|} < \infty
\end{equation}
where $\left|F\right|$ are the number of plaquettes (polygonal faces) in the 2-complex $\kappa$.

The proof goes through essentially without change in the case of the charged partition function, except in this case both
$6j$ symbols and modified $6j$ symbols that arise in vertex amplitudes along $\Gamma$ (see~\cite{ConradyPolyakov})
have to be shown to be bounded from above.  The modified $6j$ is a special case of a unitary evaluation of an $SU(2)$
spin network (without loops or trivial components), and thus a constant upper bound is provided by the work
of Abdesselam~\cite{AA09}, which follows on a conjecture given in~\cite{GVdW09} by Garoufalidis and Van der Veen.

\end{appendices}

\end{document}